\documentclass{ws-ijmpa}
\usepackage{amssymb}
\usepackage{amsmath,graphicx,color,epsfig}
\usepackage{slashed}
\begin{document}

\title{Lepton Flavor Violating Decays of $B$ and $K$ Mesons in Models with Extended Gauge Group}

\author{Fayyazuddin}

\address{National Centre for Physics, Quaid-i-Azam University Campus, Islamabad 45320, Pakistan}

\author{Muhammad Jamil Aslam}

\address{Physics Department, Quaid-i-Azam University, Islamabad 45320, Pakistan and
Institute of High Energy Physics, Chinese Academy of Sciences, Beijing 100049, China.}

\author{Cai-Dian Lu}

\address {Institute of High Energy Physics, Chinese Academy of Sciences, Beijing 100049, China and
School of Physics, University of Chinese Academy of Sciences, Beijing 100049, China.}

\maketitle

\begin{abstract}
Lepton Flavor Violating (LFV) decays are forbidden in the Standard Model (SM) and to explore them one has to go beyond it. The flavor changing neutral current induced Lepton Flavor Conserving  and LFV decays of $K$ and $B$ mesons are discussed in the gauge group $G = SU(2)_L\times U(1)_{Y_1}\times SU(2)_X$. The lepto-quark $X^{\pm 2/3}_{\mu}$ corresponding to gauge group $ SU(2)_X$ allows the quark-lepton transitions and hence giving a framework to construct the effective Lagrangian for the LFV decays. The mass of lepto-quark $m_{X}$ provides a scale at which the gauge group $G$ is broken to the SM gauge group. Using the most stringent experimental limit $\mathcal{B}(K^{0}_{L}\to \mu^{\mp}e^{\pm}) < 1.7 \times 10^{-12}$, the upper bound on the effective coupling constant $(\frac{G_{X}}{G_F})^2 < 1.1 \times 10^{-10}$ is obtained for certain pairing of lepton and quark generations in the representation $(2, \bar{2})$ of the group $G$. Later, the effective Lagrangian for the LFV meson decays for the gauge group $G = \left[SU(2)_{L}\times SU(2)_{R}\times U(1)_{Y_1^{\prime}}\right]\times SU(2)_{X}$ is constructed. Using $\mathcal{B}(K^{-} \to \pi^{-}\nu \bar{\nu}) = (1.7 \pm 1.1)\times 10^{-10}$, the bound on the ratio of effective couplings is obtained to be $\left(\frac{G_{X}}{G_{F}}\right)^2 < 10^{-10}$. A number of decay modes are discussed which provide a promising area to test this model in the current and future particle physics experiments.

\end{abstract}


\markboth{Fayyazuddin, M. J. Aslam and Cai-Dian Lu} {Lepton Flavor Violating Decays of $B$ and $K$ Mesons in Models with Extended Gauge Group}
%\catchline{}{}{}{}{}
%
%%%%%%%%%%%%%%%%%%%%%%%%%%%%%%%%%%%%%%%%%%%%%%%%%%%%%%%%%%%%%%%%%%%%

\address{National Centre for Physics, Quaid-i-Azam University Campus,
Islamabad 45320, Pakistan}

\begin{history}
\received{Day Month Year}
\revised{Day Month Year}
\end{history}

\section{Introduction}

The Standard Model is one of the most successful models of the second half of the last century. It was well established experimentally even before the discovery of the Higgs boson. Finally with the discovery of the Higgs boson with mass	$125.09\pm 0.21 (stat.) \pm 0.11 (syst.)$ \text{GeV}, in 2012 at the Large Hadron Collider (LHC) at the CERN, all of the contents of the SM are complete. Though the SM has some landmark successes, still it is not a complete theory. It is well known that the masses of all the particles are set through their interaction with the Higgs field or in some sense with Higgs boson. Compared to the Planck mass ($10^{19}$ \text{GeV}) which is the fundamental unit of mass in the theory of gravitation, the electroweak scale is orders of magnitude smaller and this is called the "hierarchy problem". The physics beyond the SM, commonly known as new physics (NP) may be revealed in the region between the electroweak unification scale \text{260} \text{GeV} and the Planck mass.

%%%%%%%%%%%%%%%%%%%%%%%%%%%%%%%%%%%%%%%%%%%%
In the Standard Model (SM), lepton and baryon numbers are conserved; $\Delta L = 0$, $\Delta B = 0$. No process with $\Delta L_{e} \neq 0$, $\Delta L_{\mu} \neq 0$ and $\Delta L_{\tau} \neq 0$ is allowed in the SM. The left-handed weak currents are coupled to the gauge vector boson $W^{+}_{\mu}$ in the SM
\begin{eqnarray}
J^{+}_{\mu}(\text{leptons}) &=& (\bar{\nu}_e,\; \bar{\nu}_{\mu},\; \bar{\nu}_{\tau})\gamma^{\mu}(1-\gamma_{5})\left(\begin{array}{c}e\\\mu\\\tau\end{array}\right)= \bar{\nu}_e \gamma^{\mu}(1-\gamma_{5})e+\cdots ,\notag \\
J^{+}_{\mu}(\text{quarks}) &=& (\bar{u},\; \bar{c},\; \bar{t})\gamma^{\mu}(1-\gamma_{5})V\left(\begin{array}{c}d\\s\\b\end{array}\right)=\bar{u}\gamma^{\mu}(1-\gamma_{5})(V_{ud}d + V_{us}s + V_{ub}b)+\cdots , \notag
\end{eqnarray}
where $V$ is the Cabibbo-Kobayashi-Maskawa (CKM) matrix. Thus unlike the quark sector, where the flavor changing weak decays are allowed, in the lepton sector the lepton flavor changing decays are forbidden. The weak neutral currents for quark level transitons are  flavor conserving in the SM and the flavor-changing-neutral-currents \text{(FCNC)}
induced weak decays are not allowed at tree level in the SM. These decays are
highly suppressed due to the Glashow-Iliopoulos-Maiani (GIM) mechanism and can only occur through loop diagrams in the SM. During last couple of years, the LHCb collaboration has measured a number of angular observables in different $B$ decays, especially in $B \to K^{*}\mu^{+}\mu^{-}$ decays. A discrepancy of $3\sigma$ is reported in the famous $P^{\prime}_5$ asymmetry observable \cite{P51,P52}. In case of $B \to K^{*}\mu^{+}\mu^{-}$ decays, there are the tensions between the measured and calculated values of decay rate in the SM \cite{Drate}. In addition, the LHCb collaboration has determined the ratio of the branching fractions of the $B^{+} \to K^{+} \mu^{+} \mu^{-}$ and  $B^{+} \to K^{+} e^{+} e^{-}$ to be $R_{K} = 0.745^{+0.090}_{-0.074}(\text{stat}) \pm 0.036(\text{syst})$ in the dilepton invariant mass squared range of $1 - 6$ GeV$^2$ \cite{LFU1,LFU2}. In the SM, there is a Lepton flavor Universality (LFU) i.e., $R_{K} = 1$, so this LHCb measurement of $R_{K}$ deviates from the SM value by $2.6\sigma$. Even the measured value of the decay rate of $B_{s}\to \phi \mu^{+}\mu^{-}$ is less than the SM predictions by $3\sigma$ \cite{Drate2}. It is worth emphasizing that all the anomalies observed in the decays induced by the FCNC transitions are lepton flavor conserving in the SM.

In the SM the neutrinos are massless and this leads to the conservation of lepton number for each generation. 
%the generation of lepton numbers is exactly conserved because of the masslessness of neutrinos in this model. 
However, the non-zero mass of neutrinos that is established through the observations of neutrino oscillation has provided evidence of the lepton number violation in the SM. In 2015, the CMS collaboration \cite{CMS} has measured the branching ratio of LFV Higgs decay $h \to \tau \mu$ to be $\mathcal{B}(h\to \tau \mu) = 0.84^{+0.39}_{-0.37}$ which deviates by $2.6\sigma$ from the SM value. This has triggered a lot of interest in the LFV decays of leptons e.g. $\ell_{i} \to \ell_{j} \gamma$, $\ell_{i} \to \ell_{j}\bar{\ell}_{k}\ell_{k}$ along with the different LFV decays of $B$ and $K$ mesons and there exists a number of new physics (physics beyond SM) studies on these decays \cite{LFVdecays}. The direct hint of the LFV decays is still missing in the experiments, however, we have upper bounds on some of these decays \cite{pdg}.

One of the possible way to reveal the new physics (NP) in between the electroweak unification scale \text{260} \text{GeV} and the Planck mass is to extend the weak unification group to a higher group,  where the extra vector boson in the extended group would provide the intermediate scale. 
In this paper, we consider the extension of group $SU(2)_L\times
U(1)_Y$ to $SU(2)_L\times U(1)_{Y_1}\times SU(2)_X$ to explore the lepton flavor violating weak decays. After formulating the model with this extended gauge group the Lagrangian for the LFV leptonic and semileptonic decays of $K$ and $B$ mesons will be worked out.  In the second part (Section 3), the extension of left-right symmetric gauge model to the gauge group $SU(2)_L\times SU_R(2)\times U^{\prime}_{Y_1}(1)\times SU(2)_X$ is considered. After developing the model, its phenomenological consequences in lepton flavor conserving and violating $K$ and $B$ decays will be discussed. Finally, the Section 4 summaries the main results of this study.

\section{Gauge Group $SU(2)_L\times U(1)_{Y_1}\times SU(2)_X$ }
The contents of gauge group $SU(2)_L\times U(1)_{Y_1}\times SU(2)_X$ are:
\begin{itemize}
\item $SU(2)_L$: The triplet of gauge bosons $(W^{\pm}_{\mu},\; W^{0}_{\mu})$ belong to the adjoint representation of $SU(2)_L$ and the corresponding gauge coupling is $g$.
\item $U(1)_{Y_1}$: $B_{1\;\mu}$ is singlet and the coupling constant is $g_{1}$.
\item $SU(2)_X$ : The gauge bosons triplet $(X_{\mu}^{\pm 2/3},\;X_{\mu}^{0})$ belong to the adjoint representation of $SU(2)_X$ and the coupling constant is  $g_X$. For lepto-quarks $Q = \left(\begin{array}{cc} 0 & 1 \\ -1 & 0 \end{array}\right) + i\tau_{2}\frac{B-L}{2} = \left(\begin{array}{cc} 0 & 2/3 \\
-2/3 & 0 
\end{array}\right)$.
\end{itemize}

The lepton $\left(\begin{array}{c} \nu_i \\ \ell_i\end{array}\right)_L$ and quark $\left(\begin{array}{c} u^{a}_i \\ d^{\prime a}_i\end{array}\right)_L$ are the doublets of  $SU(2)_L$ with hypercharge $Y_{1}=0$ and  $Y_{1}=-\frac{2}{3}$, respectively. We note that for quarks $B-L = -2/3 = Y_{1}$. The two doublets belonging to fundamental representation $\bar{2}$ of $SU(2)_X$ are $(\nu_{i},\; u^{a}_{i})_{L}$ and $(l_i,\; d^{\prime a}_{i})_{L}$, where $i$ and $a$ are the flavor the color indices, respectively. Thus, the leptons and quarks can be written in the multiplet form 
\begin{equation}
\Psi =
{
_{SU(2)_L}{\Bigg\downarrow}{}
\overset{\xleftarrow{~~SU(2)_X~~}}{\begin{pmatrix}
 \nu_i &  u_i  \\
e_i & d^{\prime}_i
\end{pmatrix}}_{L}
},
\label{matrix-sec1}
\end{equation}
where the index $a$ is suppressed and the hypercharge $Y_{1}$ is $0$ and $-\frac{2}{3}$ for the leptons and quarks, respectively. The
lepto-quarks  $X_{\mu}^{\pm 2/3}$ carry both the lepton and baryon numbers; $\Delta B = 1$, $\Delta L = -1$,  $\Delta (B+L) = 0$. The total charge $Q$ is given by
\begin{eqnarray}
Q &=& I_{3L}+I_{3X}+\frac{Y_1}{2}\notag \\
&\equiv &  I_{3L} + \frac{Y}{2},
\end{eqnarray}
with $Y = 2 I_{3X}+Y_{1}$. The hypercharge for the right handed singlets is 
\begin{equation}
Y_1 = \left \{\begin{array}{c}-2 \quad\quad\quad e_{n}, \\\quad 4/3 \quad\quad u_{n}, \\-2/3 \quad\quad d_{n}. \end{array} \right. \label{Y1values}
\end{equation}
It is to recall that only the left-handed neutrinos, $\nu_{i}$'s are present in this model.

\subsection{Interaction Lagrangian}

It is straight forward to write the Lagrangian for which the interaction
part is given by
\begin{eqnarray}
\mathcal{L}_{\text{int}}&=& -\big[\bar{\psi}_{\alpha}^{k}\gamma^{\mu}(g\vec{\tau}.\vec{W}_{\mu}+g_{1}Y_{1}B_{1\mu})_{k}^{n}\psi_{n}^{\alpha} -\bar{\psi}_{\alpha}^{k}\gamma^{\mu}g_{X}(\vec{\tau}.\vec{X}_{\mu})_{\beta}^{\alpha}\psi_{k}^{\beta}\big]\notag\\
&-&\frac{g_{1}}{2}\big[-2\bar{\ell}_{iR}\gamma^{\mu}\ell_{iR}+\frac{4}{3}\bar{u}_{iR}\gamma^{\mu}u_{iR}-\frac{2}{3}\bar{d}_{iR}\gamma^{\mu}d_{iR}\big]B_{1\; \mu} ,\label{3a}
\end{eqnarray}
where the indices $(\alpha\;, \beta)$ and $(k,\; n)$ correspond to the gauge groups $SU(2)_X$ and $SU(2)_L$, respectively. In the elaborated form, Eq. (\ref{3a}) takes the form
\begin{eqnarray}
\mathcal{L}_{\text{int}}&=&-\big[(\bar{\nu}_i,\; \bar{\ell}_{i})_{L}\gamma^{\mu}\left(\begin{array}{cc} gW^{0}_{\mu}& \sqrt{2}g W^{+}_{\mu}\\
\sqrt{2}g W^{-}_{\mu} & -gW^{0}_{\mu}\end{array}\right)\left(\begin{array}{c} \nu_i \\ \ell_i\end{array}\right)_L\notag\\
&&+ (\bar{u}_i,\; \bar{d}^{\prime}_{i})_{L}\gamma^{\mu}\left(\begin{array}{cc} gW^{0}_{\mu}-\frac{2}{3}g_1B_{1\; \mu}& \sqrt{2}g W^{+}_{\mu}\\
\sqrt{2}g W^{-}_{\mu} & -gW^{0}_{\mu}-\frac{2}{3}g_1B_{1\; \mu}\end{array}\right)\left(\begin{array}{c} u_i \\ d^{\prime}_i\end{array}\right)_L\notag \\
&&-(\bar{\nu}_i,\; \bar{u}_{i})_{L}\gamma^{\mu}\left(\begin{array}{cc} g_{X}X^{0}_{\mu}& \sqrt{2}g_{X} X^{-2/3}_{\mu}\\
\sqrt{2}g_{X} X^{2/3}_{\mu} & -g_{X}X^{0}_{\mu}\end{array}\right)\left(\begin{array}{c} \nu_i \\ u_i\end{array}\right)_L \notag\\
&&+ (\bar{\ell}_i,\; \bar{d}^{\prime}_{i})_{L}\gamma^{\mu}\left(\begin{array}{cc} g_{X}X^{0}_{\mu}& \sqrt{2}g_{X} X^{-2/3}_{\mu}\\
\sqrt{2}g_{X} X^{2/3}_{\mu} & -g_{X}X^{0}_{\mu}\end{array}\right)\left(\begin{array}{c} \ell_i \\ d^{\prime}_{i}\end{array}\right)_L\big]\notag \\
&&-\frac{g_1}{2}\big[(-2\bar{\ell}_{i R}\gamma^{\mu}\ell_{i R}+\frac{4}{3}\bar{u}_{i R}\gamma^{\mu}u_{i R}-\frac{2}{3}\bar{d}_{i R}\gamma^{\mu}d_{i R})\big]B_{1\; \mu}.\label{3}
\end{eqnarray}
From Eq. (\ref{3}), the part of Lagrangian that describe the interaction between fermions and charged gauge bosons can be written as
\begin{eqnarray}
L_{\text{int}}(\text{Charged})&=&-{g\over
2\sqrt2}\big[(\bar\nu_i\Gamma^{\mu}_L\ell_i+\bar u_i\Gamma^{\mu}_L
d_i^{\prime})W_{\mu}^{+}+h.c\big] \notag\\
&&+ {g_{X}\over
2\sqrt2}\big[\big(\bar\nu_i\Gamma^{\mu}_L
u_i+\bar \ell_i\Gamma^{\mu}_L
d^{\prime}_i\big)X_\mu^{-2/3}+h.c\big]\label{4},
\end{eqnarray}
where $\Gamma_L^\mu=\gamma^\mu(1-\gamma^5)$. Similarly, the neutral part of Lagrangian given in Eq. (\ref{3}) is
\begin{eqnarray}
L_{\text{int}}(\text{Neutral})&=&-g\sin\theta_W J^{\mu}_{em}A_{\mu}-{g\over\cos\theta_W}J_{Z}^{\mu}Z_{\mu}-\frac{g_X}{\sqrt{1-g^{\prime
2}/g_X^2}}J_{Z^{\prime}}^{\mu}Z^{\prime}_{\mu}\label{5}
\end{eqnarray}
with electromagnetic current
\begin{equation}
J_{\text{em}}^{\mu}=-\bar{\ell}_i\gamma^{\mu} \ell_i+{2\over 3}\bar
u_i\gamma^{\mu}u_i -{1\over 3}\bar
d_i\gamma^{\mu}d_i \label{6},
\end{equation}
and the weak neutral currents are
\begin{eqnarray}
J_{Z}^{\mu}&=&{1\over4}\Big[(\bar u_i\Gamma^{\mu}_L u_i+\bar
d_i\Gamma^{\mu}_Ld_i -\bar
\ell_i\Gamma^{\mu}_L\ell_i+\bar\nu_i\Gamma^{\mu}_L\nu_i-4\sin^2\theta_W J_{\text{em}}^{\mu})\Big]\notag ,\\
J_{Z^{\prime}}^{\mu}&=&{1\over4}\Big[(-\bar u_i\Gamma^{\mu}_L u_i-\bar
d_i\Gamma^{\mu}_Ld_i +\bar
\ell_i\Gamma^{\mu}_L\ell_i +\bar\nu_i\Gamma^{\mu}_L\nu_i)\notag\\
&&-{g^{\prime 2}\over g_X^{2}}\big(-4J_{\text{em}}^{\mu}+\bar
u_i\Gamma^{\mu}_L u_i-\bar d_i\Gamma^{\mu}_Ld_i
+\bar\nu_i\Gamma^{\mu}_L\nu_i-\bar
\ell_i\Gamma^{\mu}_L\ell_i\big)\Big]\label{7}.
\end{eqnarray}
In order to obtain $L_{\text{int}}(\text{Neutral})$ in the final form, we have used the following relations between neutral vector bosons $W_{\mu}^{0},\; X_{\mu}^{0}$, $B_{1\;\mu}$ and the physical neutral vector gauge bosons $A_{\mu},\; Z_{\mu}$, $Z^{\prime}_{\mu}$:
\begin{eqnarray}
gW_{\mu}^{0}&=& eA_{\mu}+g\cos \theta_{W} Z_{\mu}\notag ,\\
g_{X}X_{\mu}^{0}&=& eA_{\mu}-{g\over \cos \theta_{W}}\sin^{2}\theta_{W}Z_{\mu}-{g_{X}\over \sqrt{1-{g^{\prime 2} \over g_X^{2}}}}\big(1-{g^{\prime 2} \over g_X^{2}}\big)Z^{\prime}_{\mu}\notag ,\\
g_{1}B_{1\;\mu}&=&eA_{\mu}-{g\over \cos \theta_{W}}\sin^{2}\theta_{W}Z_{\mu}+{g_{X}\over \sqrt{1-{g^{\prime 2} \over g_X^{2}}}}Z^{\prime}_{\mu} \label{8}.
\end{eqnarray}
The electromagnetic and gauge couplings are related by
\begin{eqnarray}
{1\over g^{\prime 2}}&=&{1\over g^{2}_1}+{1\over g^{2}_X}\notag ,\\
{e\over g}&=&\sin \theta_{W}, \quad\quad\quad {e\over g^{\prime}}=\cos \theta_{W} \label{9}.
\end{eqnarray}

In order to generate the masses for differetn gauge bosons, the gauge group $SU(2)_L\times U(1)_{Y_1}\times SU(2)_X$ is
spontaneously broken to the group $SU(2)_L\times U_Y(1)$ by
introducing a scalar doublet $\eta$ that is $\bar{2}$ of $SU(2)_X$ but singlet under  $SU(2)_L$ :
\begin{eqnarray}
\eta &=&( \eta^{0}, \eta^{2/3})\notag  ,\\
 Y_1&=& ( 1, \quad {1\over 3})\notag ,\\
 Q & = & I_{3}+\frac{Y_1}{2},\quad\quad Q_{\eta^{0}} = 0, \quad\quad Q_{\eta^{2/3}} = 2/3\label{10}.
\end{eqnarray}
The interaction Lagrangian of the scalar doublet $\eta$ is
\begin{eqnarray}
\mathcal{L}_{\text{int}}(\eta) &=& -\frac{1}{2}\Big[\partial^{\mu}\eta-{i\over
2}g_X\eta\overrightarrow{\tau_X}.\overrightarrow{X}^{\mu}+{i\over
2}g_1\eta
B^{\mu}_1-{i\over3}g_1\eta^{2/3}B^{\mu}_1\Big]\notag\\
 &&\times \Big[\partial_{\mu}\bar\eta+{i\over
2}g_X\overrightarrow{\tau_X}.\overrightarrow{X}_{\mu}\bar\eta-{i\over
2}g_1
B_{1\mu}\bar\eta+{i\over3}g_1B_{1\mu}\eta^{-2/3}\Big]\label{11}.
\end{eqnarray}
For spontaneous symmetry breaking, $\eta$ can be written in the form:
\begin{align}
\eta=(\eta^{0},
\eta^{2/3})=&\left(\frac{v^{\prime}+H_X+ih_X}{\sqrt2}, \eta^{2/3}\right)\notag\\
\rightarrow&\left(\frac{v^{\prime}+H_X}{\sqrt2}, 0\right)\notag .
\end{align}
It can be noticed that the scalars $\eta^{2/3}$ and $h_X$ have been
absorbed to give masses to $X_{\mu}^{\pm 2/3}$ and $Z^{\prime}$
respectively. The mass term for the vector boson is given by
\begin{eqnarray}
\text{Mass-term}& = &-{1\over4}\left(\frac{v^{\prime}+H_X}{\sqrt2}, 0\right)\Big(g_X\overrightarrow{\tau_X}.\overrightarrow{X}^{\mu}
+g_1B_1^{\mu}\Big)\Big(g_X\overrightarrow{\tau_X}.\overrightarrow{X}_{\mu}
+g_1B_{1\mu}\Big)\left(\begin{array}{cc}
\frac{v^{\prime}+H_X}{\sqrt2}\\ 0\end{array}\right)\notag\\
& = &-{1\over4}\left(\frac{v^{\prime}+H_X}{\sqrt2}\right)^2\Big[2g_X^{2}X^{+ 2/3
\mu}X_{\mu}^{-2/3}+\frac{g_X^{2}}{1- {g^{\prime 2}\over g_X^{2}}}Z^{\prime\mu}Z^{\prime}_{\mu}\Big]\notag .
\end{eqnarray}
Hence, we have
\begin{eqnarray}
m_X^{2}={1\over4}g_X^{2}v^{\prime2}, \quad\quad\quad m^{2}_{Z^{\prime}}={1\over4}\frac{g_X^{2}v^{\prime 2}}{(1-g^{\prime
2}/g_X^2)}= \frac{m_X^{2}}{(1-g^{\prime 2}/g_X^2)}\label{8p}.
\end{eqnarray}
Thus at the mass scale $m_{X}$, the lepto-quark  $X_{\mu}^{\pm 2/3}$ and the vector boson $Z^{\prime}_\mu$ are decoupled and they acquire heavy masses that provide an intermediate mass scale, i.e., the scale between electroweak unification $(260)$ GeV and the Planck mass. The
stringent experimental limits on the LFV decays which occure through the exchange of the lepto-quark $X_{\mu}^{\pm 2/3}$ provide
lower bounds on the intermediate mass scale $m_{X}$.

\subsection{Effective Lagrangian for Lepton Flavor Violating (LFV) Decays}

From the interaction Lagrangian given in Eq. (\ref{4}), the lepton flavor violating part is
\begin{eqnarray}
\mathcal{L}(\text{LFV})={g_{X}\over
2\sqrt2}\big[\big(\bar\nu_i\Gamma^{\mu}_L
u_i+\bar \ell_i\Gamma^{\mu}_L
d^{\prime}_i\big)X_\mu^{-2/3}+h.c\big]\label{15}.
\end{eqnarray}
From above equation (\ref{15}), the effective Lagrangian relevant for the LFV decays become
\begin{eqnarray}
\mathcal{L}_{\text{eff}}&=&{g^2_{X}\over
8m^2_{X}}\big[\bar\nu_i\Gamma^{\mu}_L
u_i+\bar \ell_i\Gamma^{\mu}_L
d^{\prime}_i + \bar{u}_{i}\Gamma^{\mu}_{L}\nu_{i}+\bar{d}^{\prime}_{i}\Gamma^{\mu}_{L}\ell_{i}\big]\times \notag\\
&&\big[\bar u_j\Gamma_{\mu\; L}
\nu_j+\bar d^{\prime}_j\Gamma_{\mu\; L}
\ell_j  + \bar{\nu}_{j}\Gamma_{\mu\; L}u_{j}+\bar \ell_{j}\Gamma_{\mu\; L}{d}^{\prime}_{j}\big]\label{15a},
\end{eqnarray}
which for the case of two charged leptons $(\ell^{-}_{i} \ell^{+}_{j})$ is given by
\begin{eqnarray}
\mathcal{L}_{\text{eff}}(\ell^{-}_{i}\ell^{+}_{j})&=&\frac{g^2_X}{8m^{2}_X}\{[\bar
\ell_i\gamma^{\mu}(1-\gamma_5)d^{\prime}_i][\bar{d}^{\prime}_{j}\gamma_{\mu}(1-\gamma_5)\ell_{j}]+h.c.\}\notag \\
&=&\frac{G_X}{\sqrt{2}}\{[\bar{d}^{\prime}_{j}\gamma^{\mu}(1-\gamma_5)d^{\prime}_i][\ell_i \gamma_{\mu}(1-\gamma_5)\ell_{j}]+ h.c.\} \label{16a},
\end{eqnarray}
where $\frac{G_X}{\sqrt2}=\frac{g^2_X}{8 m^2_X}$. Using
\begin{equation}
d^{\prime}_i = V_{ip} d_p, \quad\quad d^{\prime}_j = V_{jq} d_q,\quad\quad   \label{17}
\end{equation}
and after Fierez reordering the Lagrangian (\ref{16a}) takes the form
\begin{eqnarray}
\mathcal{L}_{\text{eff}}(\ell^{-}_{i}\ell^{+}_{j}) = \frac{G_X}{\sqrt{2}}V_{ip}V^{*}_{jq}\{[\bar{d}_{q}\gamma^{\mu}(1-\gamma_5)d_p][\ell_i \gamma_{\mu}(1-\gamma_5)\ell_{j}]+h.c.\}\label{16}.
\end{eqnarray}
The effective Lagrangian for $(\bar{\ell}_{i}\;\bar{\nu}_j)$ (c.f. Eq. (\ref{15a})):
\begin{eqnarray}
\mathcal{L}_{\text{eff}}(\bar{\ell}_{i}\;\bar{\nu}_j) &=& \frac{G_X}{\sqrt2}V_{iq}\{[\bar{\ell}_{i}\gamma^{\mu}(1-\gamma_5)d_q][\bar{u}_{j}\gamma_{\mu}(1-\gamma_5)\nu_{j}]+ h.c.\}\notag \\
&=&  \frac{G_X}{\sqrt2}V_{iq}\{[\bar{u}_{j}\gamma^{\mu}(1-\gamma_5)d_q][\bar{\ell}_{i}\gamma_{\mu}(1-\gamma_5)\nu_{j}]+ h.c.\}\label{18}.
\end{eqnarray}

From Eq. (\ref{15a}), we have
\begin{equation}
\mathcal{L}_{\text{eff}}(\bar{\nu}_{i}\;\bar{\nu}_j) = \frac{G_X}{\sqrt2}\{[\bar{\nu}_{i}\gamma^{\mu}(1-\gamma_5)u_i][\bar{u}_{j}\gamma_{\mu}(1-\gamma_5)\nu_{j}]+ h.c.\}\label{18a}
\end{equation}
which after Fierz reordering takes the form
\begin{equation}
\mathcal{L}_{\text{eff}}(\bar{\nu}_{i}\;\bar{\nu}_j) = \frac{G_X}{\sqrt2}\{[\bar{u}_{i}\gamma^{\mu}(1-\gamma_5)u_i][\bar{\nu}_{j}\gamma_{\mu}(1-\gamma_5)\nu_{j}]+ h.c.\}\label{18b}
\end{equation}

To proceed further, we consider three possible choices of assignments of leptons and quarks to the representation $(2\; , \bar{2})$ of the group $SU(3)_{L}\times SU(2)_{X}$
\begin{itemize}
\item[(i)] (1\;, 2\;, 3) \hspace{1cm} $\left(\begin{array}{cc} \nu_{e} & u \\
e & d^{\prime}\end{array}\right)_{L}$\quad, \quad\quad\quad $\left(\begin{array}{cc} \nu_{\mu} & c \\
\mu & s^{\prime}\end{array}\right)_{L}$\quad, \quad\quad\quad $\left(\begin{array}{cc} \nu_{\tau} & t \\
\tau & b^{\prime}\end{array}\right)_{L}$.\\
\item[(ii)] (1\;, 3\;, 2) \hspace{1cm} $\left(\begin{array}{cc} \nu_{e} & u \\
e & d^{\prime}\end{array}\right)_{L}$\quad, \quad\quad\quad $\left(\begin{array}{cc} \nu_{\tau} & c \\
\tau & s^{\prime}\end{array}\right)_{L}$\quad, \quad\quad\quad $\left(\begin{array}{cc} \nu_{\mu} & t \\
\mu & b^{\prime}\end{array}\right)_{L}$,\\
\item[(ii)] (2\;, 3\;, 1) \hspace{1cm} $\left(\begin{array}{cc} \nu_{\mu} & u \\
\mu & d^{\prime}\end{array}\right)_{L}$\quad, \quad\quad\quad $\left(\begin{array}{cc} \nu_{\tau} & c \\
\tau & s^{\prime}\end{array}\right)_{L}$\quad, \quad\quad\quad $\left(\begin{array}{cc} \nu_{e} & t \\
e & b^{\prime}\end{array}\right)_{L}$,
\end{itemize}
For the assignment (i), the leptonic and semileptonic $K$ decays of interest are
\begin{eqnarray}
\bar{K}^{0} \to \mu^{-}e^{+}\; : V_{cs}V^{*}_{ud}\; , \quad\quad \bar{K}^{0} \to e^{-}e^{+}\; : V_{cd}V^{*}_{ud} , \quad\quad \bar{K}^{0} \to \mu^{-}\mu^{+}\; : V_{cs}V^{*}_{us} ,\notag\\
\quad\quad K^{-} \to \pi^{-}\mu^{-}e^{+}\; : V_{cs}V^{*}_{ud} , \quad\quad K^{-} \to \pi^{-}e^{-}e^{+}\; : V_{cd}V^{*}_{ud} , \quad\quad K^{-} \to \pi^{-}\mu^{-}\mu^{+}\; : V_{cs}V^{*}_{us} \notag\\
\label{Kd1-1},
\end{eqnarray}
and of the $B$-meson are
\begin{eqnarray}
\bar{B}_{s}^{0} \to \tau^{-}\mu^{+}\; : V_{tb}V^{*}_{cs}\; , \bar{B}^{0} \to \tau^{-}e^{+}\; : V_{tb}V^{*}_{ud}\; ,  \bar{B}_{s}^{0} \to \mu^{-}\mu^{+}\; : V_{ts}V^{*}_{cs} \quad \;, \notag\\
\bar{B}^{0,\; -}  \to  (K^{0,\; -})^{*}\mu^{-}\tau^{+} \; : \quad \ \ V_{tb}V^{*}_{cs} ,
\bar{B}^{0\; -} \to  \rho^{0\; -}(\pi^{0}\; , \pi^{-} )\tau^{-}e^{+} \; : \quad \ V_{tb}V^{*}_{ud} ,   \notag \\
\quad\quad \bar{B}_{s}^{0} \to \phi \tau^{-}\mu^{+} : \quad \ V_{tb}V^{*}_{cs} , \quad \quad \bar{B}_{s}^{0} \to (\bar{K}^{0})^{*} \tau^{-}e^{+} : \quad \ V_{tb}V^{*}_{ud}\notag \\
\bar{B}_{c}  \to   (D_{s}^{-})^{*}\tau^{-}\mu^{+} \; : \quad\quad\quad \ V_{tb}V^{*}_{cs} \; , \quad\quad \bar{B}_{c}  \to   (D^{-})^{*}\tau^{-}e^{+} \; : \quad\quad V_{tb}V^{*}_{ud} \notag\\
\label{Bd1-1}.
\end{eqnarray}

For assignment (ii), the Cabbibo allowed decays of interest are
\begin{eqnarray}
\bar{K}^{0} \to e^{-}e^{+}\; : V_{us}V^{*}_{ud}\; , \quad\quad \bar{K}^{0} \to \mu^{-}e^{+}\; : V_{ts}V^{*}_{ud} ,
\notag\\
\quad\quad K^{-} \to \pi^{-}e^{-}e^{+}\; : V_{us}V^{*}_{ud} ,\quad \quad K^{-} \to \pi^{-}\mu^{-}e^{+}\; : V_{ts}V^{*}_{ud} \label{Kd2-1},
\end{eqnarray}
\begin{eqnarray}
\bar{B}_{s}^{0} \to \mu^{-}\mu^{+}\; : V_{tb}V^{*}_{ts}\; , \bar{B}_{s}^{0} \to \mu^{-}e^{+}\; : V_{tb}V^{*}_{us}\; ,  \bar{B}_{s}^{0} \to \mu^{-}\tau^{+}\; : V_{tb}V^{*}_{cs}\; ,\notag\\
 \bar{B}^{0} \to \mu^{-}e^{+}\; : V_{tb}V^{*}_{ud}\; , \bar{B}^{0} \to \mu^{-}\tau^{+}\; : V_{tb}V^{*}_{cd} \label{Bd2-1}
\end{eqnarray}
\begin{eqnarray}
\bar{B}_{s}^{0} \to \phi \mu^{-}\mu^{+}\; : V_{tb}V^{*}_{ts}\; \quad \bar{B}_{s}^{0} \to \phi \mu^{-}e^{+}\; : V_{tb}V^{*}_{us}\;, \quad \bar{B}_{s}^{0} \to \phi \mu^{-}\tau^{+}\; : V_{tb}V^{*}_{cs}\; ,\notag\\
\bar{B}^{0\;, -} \to (\bar{K}^{0\;, -})^{*}\mu^{-}\mu^{+}\; : V_{tb}V^{*}_{ts}\; ,
\bar{B}^{0\;, -} \to (\bar{K}^{0\;, -})^{*}\mu^{-}e^{+}\; : V_{tb}V^{*}_{ud}\; ,\notag \\
\quad \bar{B}^{0\;, -} \to (\bar{K}^{0\;, -})^{*}\mu^{-}\tau^{+}\; : V_{tb}V^{*}_{cs}\;, 
 B_{c}^{ -} \to (D_{s}^{-})^{*}\mu^{-}e^{+}\; : V_{tb}V^{*}_{us}\; ,\notag \\\quad B_{c}^{-} \to (D_{s}^{-})^{*}\mu^{-}\tau^{+}\; : V_{tb}V^{*}_{cs}\;, 
 \quad \Lambda_{b} \to \Lambda \mu^{-}\tau^{+}\; :  V_{tb}V^{*}_{cs} \;,  \quad\Lambda_{b} \to \Lambda \mu^{-}e^{+}\; :  V_{tb}V^{*}_{us} \notag\\
 \label{Bcs-1}
\end{eqnarray}

Similarly, for the assignment (iii), interchange $\mu$ and $e$ in Eqs. (\ref{Kd2-1}, \ref{Bd2-1}) and Eq. (\ref{Bcs-1}).
%\begin{equation}
%\bar{K}^{0} \to \mu^{-}\mu^{+}\; : V_{us}V^{*}_{ud}\; , \quad\quad \bar{K}%^{0} \to e^{-}\mu^{+}\; : V_{ts}V^{*}_{ud} ,
%\quad\quad K^{-} \to \pi^{-}\mu^{-}\mu^{+}\; : V_{us}V^{*}_{ud} , \quad\quad %K^{-} \to \pi^{-}e^{-}\mu^{+}\; : V_{ts}V^{*}_{ud} \label{Kd3-1},
%\end{equation}
%\begin{equation}
%\bar{B}_{s}^{0} \to e^{-}\mu^{+}\; : V_{tb}V^{*}_{us}\; , \bar{B}_{s}^{0} \to %e^{-}\tau^{+}\; : V_{tb}V^{*}_{cs}\; ,  \bar{B}^{0} \to e^{-}\mu^{+}\; : V_{tb}%V^{*}_{ud}\; , \bar{B}^{0} \to e^{-}\tau^{+}\; : V_{tb}V^{*}_{cd}\; , \bar{B}%_{s}^{0} \to e^{-}e^{+}\; : V_{tb}V^{*}_{ts} \label{Bd3-1}
%\end{equation}
%\begin{eqnarray}
%\bar{B}_{s}^{0} \to \phi e^{-}e^{+}\; : V_{tb}V^{*}_{ts}\; \quad \bar{B}_{s}%^{0} \to \phi e^{-}e^{+}\; : V_{tb}V^{*}_{ts}\;, \quad \bar{B}_{s}^{0} \to \phi %e^{-}\mu^{+}\; : V_{tb}V^{*}_{us}\; ,\quad \bar{B}_{s}^{0} \to \phi e^{-}%\tau^{+}\; : V_{tb}V^{*}_{cs}\;, \bar{B}^{0\;, -} \to (\bar{K}^{0\;, -})^{*}e^{-}%e^{+}\; : V_{tb}V^{*}_{ts}\; ,\notag \\
%\bar{B}^{0\;, -} \to (\bar{K}^{0\;, -})^{*}e^{-}\mu^{+}\; : V_{tb}V^{*}_{us}\; ,\quad \bar{B}^{0\;, -} \to (\bar{K}^{0\;, -})^{*}e^{-}\tau^{+}\; : V_{tb}V^{*}_{cs}\;, 
% B_{c}^{ -} \to (D_{s}^{-})^{*}e^{-}\mu^{+}\; : V_{tb}V^{*}_{us}\; ,\notag \\%\quad B_{c}^{-} \to (D_{s}^{-})^{*}\mu^{-}\tau^{+}\; : V_{tb}V^{*}_{cs}\;, 
% \quad \Lambda_{b} \to \Lambda e^{-}\mu^{+}\; :  V_{tb}V^{*}_{us} \;,  \quad%\Lambda_{b} \to \Lambda e^{-}\tau^{+}\; :  V_{tb}V^{*}_{cs} \label{Bcs-1}
%\end{eqnarray}
\subsection{Weak decays of $K$ and $B$ mesons into final state two charged lepton pair ($\ell^{-}_i\ell^{+}_j)$}

The lepton flavor violating decays of interest are the weak decays of mesons involving two charged leptons in the final state i.e., 
\begin{itemize}
\item $\bar{P}\to \ell^{-}_i\ell^{+}_j$\;, \quad\quad\quad\quad\quad $P = K_{L}^{0}\;,  B^{0}\;,  B_{s}^{0}$
\end{itemize}
%and the second are the semi-leptonic decays
%$\begin{itemize}
%\item $\bar{P}\to (\bar{P}^{\prime})^{\ast}\ell^{-}_i\ell^{+}_j$
%\end{itemize}
%where $P^{\prime}$ is final state pseudo-scalar meson.

From Eq. (\ref{16}), the branching ratio for $\bar{P}\to \ell^{-}_i\ell^{+}_j$ decays is given by 
\begin{eqnarray}
\mathcal{B}(\bar{P}\to \ell^{-}_i\ell^{+}_j)&=&
\frac{\tau_{P}}{\hbar}\left({G_X\over G_F}\right)^2 |V_{ip}V^{*}_{jq}|^2\left(1-{m^2
_{i}+m^2
_{j}\over m^2_P}\right)\left(1-{(m^2
_{i}-m^2
_{j})^2\over {m^2_P(m^2
_{i}+m^2
_{j})}}\right)\notag\\
&&\times \Big[{G^2_F\over 8\pi}f^2_P
m_P(m^2
_{i}+m^2
_{j})\Big]\label{21},
\end{eqnarray}
where $\tau_{P}$ is the decay time of initial state meson and $G_{F}$ is the Fermi coupling constant. From above equation, in the limit ${m^{2}_{e}\over m^{2}_{\mu}} << 1$, for the assignment (i) we have 
\begin{eqnarray}
\mathcal{B}(K^{0}_{L}\rightarrow e^{\pm}\mu^{\mp}) &=& 
\frac{\tau_{K^0_L}}{\hbar}\left({G_X\over G_F}\right)^2 |V_{cs}|^2|V_{ud}|^2\left(1-{m^2
_{\mu}\over m^2_K}\right)^2{1\over 2}\Big[{G^2_F\over 8\pi}f^2_K
m^2_{\mu}m_K\Big]\label{22},\\
\mathcal{B}(K^{0}_{L}\rightarrow \mu^{-}\mu^{+}) &=& 
\frac{\tau_{K^0_L}}{\hbar}\left({G_X\over G_F}\right)^2 |V_{cs}|^2|V_{ud}|^2\left(1-{2m^2
_{\mu}\over m^2_K}\right)^2 2 \Big[{G^2_F\over 8\pi}f^2_K
2 m^2_{\mu}m_K\Big]\label{23}.
\end{eqnarray}
The branching ratio for $K^{-} \to \mu^{-}\bar{\nu}_{\mu}$ in the SM can be expressed as
\begin{equation}
\mathcal{B}(K^{-} \to \mu^{-}\bar{\nu}_{\mu}) = \frac{\tau_{K^{-}}}{\hbar}\frac{G^{2}_{F}}{8\pi}|V_{us}|^2f^{2}_{K}m_{K^{-}}m^{2}_{\mu}\left(1-{m^2
_{\mu}\over m^2_K}\right)^2\label{KSM}
\end{equation}
\begin{eqnarray}
\mathcal{B}(\bar{B}^{0}_{s}\rightarrow \tau^{-}\mu^{+}) &=& 
\frac{\tau_{\bar{B}^{0}_{s}}}{\hbar}\left({G_X\over G_F}\right)^2 |V_{cs}|^2|V_{tb}|^2\left(1-{m^2
_{\tau}+m^2
_{\mu}\over m^2_{B_s}}\right)\notag\\
&&\times \left(1-{(m^2
_{\tau}-m^2
_{\mu})^2\over {m^2_{B_s}}(m^2
_{\tau}+m^2
_{\mu})}\right) \Big[{G^2_F\over 8\pi}f^2_{B_s}
 (m^2_{\tau}+m^2_{\mu})m_{B_s}\Big]\notag \\
 &\approx & \frac{\tau_{B^0_s}}{\hbar}\left({G_X\over G_F}\right)^2 |V_{cs}|^2|V_{tb}|^2\left(1-{m^2
_{\tau}\over m^2_{B_s}}\right)^2 \Big[{G^2_F\over 8\pi}f^2_{B_s}
 m^2_{\tau}m_{B_s}\Big] \label{23a}.
\end{eqnarray}
In Eq. (\ref{23a}), we have ignored the term proportional to ${m^{2}_{\mu}\over m^{2}_{\tau}}$ that is much less than one. From Eq. (\ref{23a}), the branching ratio for the decay $\bar{B}^{0}_{s}\rightarrow \mu^{-}\mu^{+}$ can be written as
\begin{equation}
\mathcal{B}(\bar{B}^{0}_{s}\rightarrow \mu^{-}\mu^{+})=\frac{\tau_{\bar{B}^0_s}}{\hbar}\left({G_X\over G_F}\right)^2 |V_{cs}|^2|V_{ts}|^2 \left(1-{2m^2
_{\mu}\over {m^2_{B_s}}}\right) \Big[{G^2_F\over 8\pi}f^2_{B_s} 2m^2_{\mu}m_{B_s}\Big]\label{24}.
\end{equation}
The corresponding branching ratio of $B_{s}\rightarrow \mu^{-}\mu^{+}$ in the Standard Model is expressed as
\begin{equation}
\mathcal{B}(\bar{B}^{0}_{s}\rightarrow \mu^{-}\mu^{+}) = \frac{\tau_{\bar{B}^0_s}}{\hbar}|V_{ts}|^2|C_{10}|^2\frac{1}{\pi}\left(1-{2m^2
_{\mu}\over {m^2_{B_s}}}\right) \Big[{G^2_F\over 16\pi}f^2_{B_s} \alpha^2 m^2_{\mu}m_{B_s}\Big]\label{25}.
\end{equation}
%Neglecting the term ${m^{2}_{\mu}\over m^{2}_{\tau}} << 1$, we have
%\begin{equation}
%\mathcal{B}(\bar{B}^{0}_{s}\rightarrow \tau^{-}\mu^{+}) = 
%\frac{\tau_{B^0_s}}{\hbar}\left({G_X\over G_F}\right)^2 |V_{cs}|^2|V_{tb}|^2\left(1-{m^2
%_{\tau}\over m^2_{B_s}}\right)^2 \Big[{G^2_F\over 8\pi}f^2_{B_s}
 %m^2_{\tau}m_{B_s}\Big] \label{26}.
%\end{equation}

\begin{table}
\caption{Numerical values of the different input parameters corresponding to the SM\protect\cite{pdg}.}\label{SM-values}
\begin{tabular}{llllll}
%{|c|c|c|c|c|c|}
	\hline\hline
 $G_F$ \ \ \ \ \  & $1.16638\times 10^{-5}$ GeV$^{-2}$& $\tau_{K^{0}_L}$ \ \ \ &  \ \ \ \ $5.116\times 10^{-8}$ s \ \ \ &  $f_{K}$ \ \ \ \ & \ \ \ \ \ 160 MeV \ \ \ \ \  \\ 
	 	 $m_{K^{0}}$ & 497.6 MeV & $m_{\mu}$ & 105.66 MeV & $V_{us} \approx V_{cd}$ & $0.223$   \\ 
	 $\tau_{B_{s}}$ & $1.512\times 10^{-12}$ s & $m_{B_{s}}$ & 5.367 GeV  &$m_{\tau}$ & 1.177 GeV \\ 
	 $f_{B_s}$ & 227.7 MeV & $V_{ts}$ & 0.04  & $C_{10}$ & 4.13  \\
	  $\alpha(m_{Z}) = \alpha$ & $1/128 $ & & & &  \\
	\hline\hline
\end{tabular}%
\end{table}

Using the values tabulated in Table \ref{SM-values} in 
 Eqs. (\ref{22}) and (\ref{23}), we have
\begin{eqnarray}
\mathcal{B}(K^{0}_{L}\rightarrow e^{\pm}\mu^{\mp}) &\approx& (27.3) \left({G_X\over G_F}\right)^2  \label{27},\\
\mathcal{B}(K^{0}_{L}\rightarrow \mu^{-}\mu^{+})&\approx&  (10.8) \left({G_X\over G_F}\right)^2 \label{28}.
\end{eqnarray}
The SM value of the branching ratio of $\bar{B}^{0}_{s}\rightarrow \mu^{-}\mu^{+}$ is
\begin{equation}
\mathcal{B}(\bar{B}^{0}_{s}\rightarrow \mu^{-}\mu^{+})_{SM} \approx 3.26 \times 10^{-9}\label{29}, 
\end{equation}
which gives
\begin{eqnarray}
\frac{\mathcal{B}(\bar{B}^{0}_{s}\rightarrow \tau^{-}\mu^{+})}{\mathcal{B}(\bar{B}^{0}_{s}\rightarrow \mu^{-}\mu^{+})_{SM} } &\approx & \left({G_X\over G_F}\right)^2 (2.64) \times 10^{9} \label{30},\\
\mathcal{B}(\bar{B}^{0}_{s}\rightarrow \tau^{-}\mu^{+})  &\approx & (8.60)\left({G_X\over G_F}\right)^2 \label{31} ,\\
\frac{\mathcal{B}(\bar{B}^{0}\rightarrow \tau^{-}e^{+})}{\mathcal{B}(\bar{B}^{0}_{s}\rightarrow \tau^{-}\mu^{+})} &\approx & \frac{\tau_{B^0_d}}{\tau_{B^0_s}}\frac{|V_{tb}|^2|V_{ud}|^2}{|V_{tb}|^2|V_{cs}|^2} \frac{f_{B^0}m_{B^0}}{f_{B^0_s}m_{B^0_s}} \approx 1\label{31-a}.
\end{eqnarray}
The most stringent experimental limit on the LFV decay is
\begin{equation}
\mathcal{B}(K^{0}_{L}\rightarrow e^{\pm}\mu^{\mp}) <4.7\times 10^{-12}\label{K-bound}.
\end{equation}
Hence using the bound given in Eq. (\ref{K-bound}), from Eq. (\ref{27}) we obtain
\begin{equation}
 \left({G_X\over G_F}\right)^2 <1.7\times 10^{-13},\quad\quad \left({G_X\over G_F}\right) <4.1\times 10^{-7}, \quad\quad\quad  \left({g\over g_X}\right) \left({m_X\over m_W}\right)>1.5 \times 10^{3}.
\end{equation}
In the symmetry limit $g = g_{X}$, the lower limit on the mass of leptoquark is $1.5 \times 10^{3} m_{W} = 120$ TeV which is too high for the experimental observation. 
Similarly from Eq. (\ref{28}, \ref{29}) and Eq. (\ref{31})
\begin{eqnarray}
\mathcal{B}(K^{0}_{L}\rightarrow \mu^{-}\mu^{+}) < 6.7\times 10^{-14}\notag ,\\
\mathcal{B}(\bar{B}^{0}_{s}\rightarrow \tau^{-}\mu^{+}) <  1.5\times 10^{-12}\notag ,\\
\mathcal{B}(\bar{B}^{0}\rightarrow \tau^{-}e^{+})  <  1.5\times 10^{-12} \label{32}.
\end{eqnarray}
We note that
\begin{eqnarray}
\mathcal{B}(B^{0}\rightarrow \tau^{+}e^{-})  &=& \mathcal{B}(\bar{B}^{0}\rightarrow \tau^{-}e^{+})\notag ,\\
 \mathcal{B}(B^{0}_{s}\rightarrow \tau^{+}\mu^{-})  &=& \mathcal{B}(\bar{B}^{0}_{s}\rightarrow \tau^{-}\mu^{+})\label{33},
\end{eqnarray}
\begin{eqnarray}
\mathcal{B}(B^{0}\rightarrow \tau^{-}e^{+})  &=& \frac{|V_{ub}|^2|V_{td}|^2}{|V_{tb}|^2|V_{ud}|^2}\mathcal{B}(B^{0}\rightarrow \tau^{+}e^{-}) <<  \mathcal{B}(B^{0}\rightarrow \tau^{+}e^{-})\notag ,\\
\mathcal{B}(B^{0}_{s}\rightarrow \tau^{-}\mu^{+})  &=& \frac{|V_{cb}|^2|V_{ts}|^2}{|V_{tb}|^2|V_{cd}|^2}\mathcal{B}(B^{0}_{s}\rightarrow \tau^{+}\mu^{-}) <<  \mathcal{B}(B^{0}_{s}\rightarrow \tau^{+}\mu^{-})\label{34}.
\end{eqnarray}
The time integrated decay rate due to quantum interference of $B^{0}$ and $\bar{B}^{0}$ for the LFV decays of $B^{0}$ to lepton pairs is a promising area to test the model based on the extended electroweak unification group considered in this paper. Writing
\begin{eqnarray}
\Gamma_{f}(t)&=&{1\over2}e^{-\Gamma
t}|A_{f}|^2|(1+\cos\Delta mt), \quad\quad A_{\bar{f}}=0\notag,\\
\Gamma_{\bar f}(t)&=&{1\over2}e^{-\Gamma
t}|\bar{A}_{\bar{f}}|^2(1-\cos\Delta mt)\notag.
\end{eqnarray}
The $CP$-invariance implies $\bar{A}_{\bar{f}} = A_{f}$,
\begin{eqnarray}
\frac{\int^{\infty}_{0}\Gamma(B^{0}_{s}\rightarrow \tau^{-}\mu^{+})dt}{\int^{\infty}_{0}[\Gamma(B^{0}_{s}\rightarrow \tau^{+}\mu^{-})+\Gamma(B^{0}_{s}\rightarrow \tau^{-}\mu^{+})]dt} &=& \frac{\chi^2_s}{2(1+\chi^{2}_s)} = \chi_{s} \approx 0.5\notag ,\\
\frac{\int^{\infty}_{0}\Gamma(B^{0}\rightarrow \tau^{-}e^{+})dt}{\int^{\infty}_{0}[\Gamma(B^{0}\rightarrow \tau^{+}e^{-})+\Gamma(B^{0}\rightarrow \tau^{-}e^{+})]dt} &=& \chi_{d} \approx 0.19\label{35}.
\end{eqnarray}
Hence for time integrated decay of  $B_{s}^{0}$ into lepton pairs, we have
\begin{equation}
\mathcal{B}(B^{0}_{s}\rightarrow \tau^{-}\mu^{+}) = \mathcal{B}(B^{0}_{s}\rightarrow \tau^{+}\mu^{-}),
\end{equation}
although $\mathcal{B}(B^{0}_{s}\rightarrow \tau^{-}\mu^{+}) << \mathcal{B}(B^{0}_{s}\rightarrow \tau^{+}\mu^{-})$.
Thus for the time integrated decay of $B_{s}^{0}$, there are equal pairs of $(\tau^{-}\mu^{+})$ and $(\tau^{+}\mu^{-})$. 

For the assignment $(ii)$: Using Eq. (\ref{Kd2-1}) along with Eq. (\ref{24}) and Eq. (\ref{31}), we obtain
\begin{equation}
\mathcal{B}(K^{0}_{L}\rightarrow e^{\pm}\mu^{\mp}) \approx (27.3) \left({G_X\over G_F}\right)^2|V_{ts}|^2 \label{Bound-K2},
\end{equation}
which gives $\left({G_X\over G_F}\right)^2 < 1.1 \times 10^{-10}$. We will get the same value if we consider the assignment $(iii)$. For these cases the limit on the mass of lepto-quark is $m_{X}\geq 25$TeV. This limit will be more refined, once we will have more control on the experimental observations of flavor violating $B$ decays.

Hence for experimental detection of LFV decays of $B$ mesons, the assignments $(ii)$ and $(iii)$ are suitable. In assignment $(ii)$, the predicted values of branching ratio for leptonic $B$-meson decays given in Eqs. (\ref{Bd1-1}) from Eq. (\ref{30}) and using  $\left({G_X\over G_F}\right)^2 < 1.1 \times 10^{-10}$ are
\begin{eqnarray}
\mathcal{B}(\bar{B}_{s}^{0}  \to  \mu^{-}\tau^{+}) & < & 9.5\times 10^{-10} \notag ,\\
\mathcal{B}(\bar{B}^{0}  \to  \mu^{-}\tau^{+}) & < & 4.5 \times 10^{-11} \notag ,\\
\mathcal{B}(\bar{B}^{0}  \to  \mu^{-}e^{+})  & = &\frac{9.5 \times 10^{-10}}{\left(1-m_{\tau}^2/m^{2}_{B}\right)^2}\frac{m^2_{\mu}}{m^2_{\tau}} < 4.3\times 10^{-12} \notag .
\end{eqnarray}
For the assignment $(iii)$, we just have to interchange $\mu$ and $e$.

\subsection{Semi-leptonic flavor violating decays}

In this section, we discuss the decays of the type $\bar{P} \to (\bar{P}^{\prime})\ell^{-}_{1}\ell^{+}_2$, where $\ell$ can be $e,\; \mu$ and $\tau$ leptons and $\bar{P}^{(\prime)}$ correspond to the pseudo-scalar meson, for the assignments $(ii)$ and $(iii)$. The $T$ matrix in this case is given by
\begin{equation}
T = -\frac{G}{\sqrt{2}}\frac{1}{(2\pi)^3}\sqrt{\frac{m_{1}m_{2}}{k_{1\;0}k_{2\;0}}}\langle \bar{P}^{\prime}(p^{\prime})|\bar{q}\gamma^{\lambda}Q|\bar{P}(p)\rangle (\bar{u}(k_{1})\gamma_{\lambda}(1-\gamma_{5})v(k_2)) \label{sLFV1},
\end{equation}
where $G = G_{F}V_{tb}(V^{*}_{cs},\; V^{*}_{us})$. The matrix element $\langle \bar{P}^{\prime}(p^{\prime})|\bar{q}\gamma^{\lambda}Q|\bar{P}(p)\rangle$ can be parameterized in terms of the form factor $f_{+}(t)$ and $f_{-}(t)$ as
\begin{eqnarray}
\langle \bar{P}^{\prime}(p^{\prime})|\bar{q}\gamma^{\lambda}Q|\bar{P}(p)\rangle &=&\frac{1}{(2\pi)^3}\frac{1}{\sqrt{4p_{0}p^{\prime}_{0}}}\left[f_{+}(t)(p+p^{\prime})^{\lambda}+f_{-}(t)(p-p^{\prime})^{\lambda}\right]\notag \\
&=&\frac{1}{(2\pi)^3}\frac{1}{\sqrt{4p_{0}p^{\prime}_{0}}}\big[f_{+}(t)\left((p+p^{\prime})^{\lambda}-\frac{m^2_{P}-m^2_{P^{\prime}}}{t}q^{\lambda}\right)\notag\\
&&+f_{0}(t)\frac{m^2_{P}-m^2_{P^{\prime}}}{t}q^{\lambda}\big]\label{sLFV2},
\end{eqnarray}
where
\begin{equation}
f_{0}(t) = f_{+}(t)+\frac{t}{m^2_{P}-m^2_{P^{\prime}}}f_{-}(t)\label{sLFV3},
\end{equation}
with $q = (p-p^{\prime}) = (k_{1}+k_{2})$ and $t = q^2$.

In case of $m_1 = m_2 = m_{\ell}$, from Eq. (\ref{sLFV8}), one obtains
\begin{equation}
\frac{d\Gamma}{dt} = \frac{G^2}{24\pi^3}K\sqrt{1-\frac{4m^{2}_{\ell}}{t}}\left[K^2\left(1-\frac{m^2_{\ell}}{t}\right)f^2_{+}(t)+\frac{3}{4}\frac{(m^2_{P}-m^2_{P^{\prime}})^2}{m^2_{P}}\frac{m^2_{\ell}}{t}f^2_{0}(t)\right]\label{sLFV9}.
\end{equation}

It has to be noticed from Eq. (\ref{sLFV4}) that $t$ is maximum, when $\omega$ is minimum. The minimum value of $\omega$ is $m_{P^{\prime}}$. Thus, the limits of integration on $t$ becomes, $t_{\text{min}} = 4m^2_{\ell}$, $t_{\text{max}} = m^2_{P}\left(1-\frac{m^2_{P^{\prime}}}{m^2_{P}}\right)$. In particular for the decay $K^{-} \to \pi^{-}e^{+}e^{-}$, $G = G_{F}|V_{us}V^{*}_{ud}|$, $m_{P}$, $m_{P^{\prime}}$ and $m_{\ell}$ are the masses of $K$, $\pi$ and $e$, respectively.

In case of $B^{-} \to K^{-}\mu^{+}\tau^{-}$ decay, and working in the limit $m^2_{\mu}/m^2_{\tau} \approx 0$, the expression for the differential decay rate becomes (for details see Appendix)
\begin{equation}
\frac{d\Gamma}{dt} = \frac{G_{F}^2}{24\pi^3}|V_{tb}V^{*}_{cs}|^2K\left(1-\frac{m_{\tau}^{2}}{t}\right)^2\left[K^2\left(1+\frac{m^2_{\tau}}{2t}\right)f^2_{+}(t)+\frac{3}{8}\frac{(m^2_{B}-m^2_{K})^2}{m^2_{B}}\frac{m^2_{\tau}}{t}f^2_{0}(t)\right]\label{sLFV10}
\end{equation}
and the limits of integration in this particular case are $t_{\text{min}} = m^2_{\tau}$, $t_{\text{max}} = m^2_{B}\left(1-\frac{m^2_{K}}{m^2_{B}}\right)$.

Similarly for the decay $B^{-} \to K^{-}\mu^{-}e^{+}$ and working in the limit $m^2_{e}/m^2_{\mu} \approx 0$, in Eq. (\ref{sLFV10}), we have to replace $m_{\tau}$ by $m_{\mu}$ and $|V_{tb}V^{*}_{cs}|$ with $|V_{tb}V^{*}_{us}|$. In case of the assignment $(iii)$, we just have to interachange $\mu$ and $e$ in the final state.

\section{Gauge Group: $\left[SU(2)_{L}\times SU(2)_{R}\times U(1)_{Y_1^{\prime}}\right]\times SU(2)_{X}$}
The contents corresponding to the gauge group $\left[SU(2)_{L}\times SU(2)_{R}\times U(1)_{Y_1^{\prime}}\right]\times SU(2)_{X}$ are the following:
\begin{itemize}
\item $SU(2)_{L,R}$: The triplet of gauge bosons $(W^{\pm}_{L, R\; \mu},\; W^{0}_{L, R\; \mu})$ belonging to adjoint representation of $SU(2)_{L, R}$ and the corresponding gauge coupling is $g$.
%SU_R(2)&:& (W^{\pm}_{R \mu},\; W^{0}_{R\; \mu}): \text{the triplet belonging %to adjoint representation} SU(2)_L; g\notag \\
\item $U(1)_{Y^{\prime}_1}$: $B^{\prime}_{1\;\mu}$, singlet and the coupling constant is $g^{\prime}_{1}$.
\item $SU(2)_X$: The triplet $(X_{\mu}^{\pm 1/3},\;X_{\mu}^{0})$ belong to the adjoint representation of $SU(2)_X$ and the corresponding coupling constant is $g_X$.
\end{itemize}
Leptons and quarks are assigned to the multiplets
%\begin{eqnarray}
%\Psi = \left(\begin{array}{cc} \nu_i & d^{\prime c}_i \\
%e_i & -u^{\prime c}_i\end{array}\right)_L,\notag\\
%\end{eqnarray}
\begin{equation}
\Psi =
{
_{SU(2)_L}{\Bigg\downarrow}{}
\overset{\xleftarrow{~~SU(2)_X~~}}{\begin{pmatrix}
 \nu_i & d^{\prime c}_i \\ 
e_i & -u^{\prime c}_i
\end{pmatrix}}_{L} \Bigg\downarrow_{ SU(2)_R}
}
\label{matrix1-1}
\end{equation}
and
\begin{equation}
\Psi^{\prime} =
{
_{SU(2)_L}{\Bigg\downarrow}{}
\overset{\xleftarrow{~~SU(2)_X~~}}{\begin{pmatrix}
 u_i & e^{c}_i \\ 
d^{\prime c}_i  & -N^{c}_i
\end{pmatrix}}_{L} \Bigg\downarrow_{ SU(2)_R}
}
\label{matrix1-2},
\end{equation}
respectively. The charge $Q$ is
\begin{equation}
Q = I_{3L}+I_{3R}+\frac{Y^{\prime}_1}{2}+I_{3X}.
\end{equation}
The hyper charge $Y^{\prime}_{1} = 0$ for leptons and anti-leptons, and  $Y^{\prime}_{1} = \pm 4/3$ for quarks and anti-quarks.
The lepto-quark $X^{\pm 1/3}_{\mu}$ has  $\Delta B= -1, \; \Delta L = -1$ but  $\Delta (B-L)=0$. The Interaction Lagrangian is
\begin{eqnarray}
\mathcal{L}_{\text{int}}&=&-\frac{1}{2}\left[(\bar{\nu}_i,\; \bar{\ell}_{i})_{L}\gamma^{\mu}(g\vec{\tau}\cdot  \vec{W}_{L\mu})\left(\begin{array}{c} \nu_i \\ {\ell_i}\end{array}\right)_L + (\bar{u}_i,\; \bar{d}^{\prime}_{i})_{L}\gamma^{\mu}\left(g\vec{\tau}\cdot \vec{W}_{L\mu}+\frac{4}{3}g^{\prime}_{1}B^{\prime}_{\mu}\right)\left(\begin{array}{c} u_i \\ d^{\prime}_i\end{array}\right)_L\right.\notag \\
&&+(\bar{\ell}^{c}_i,\; -N^{c}_{i})_{L}\gamma^{\mu}(g\vec{\tau}\cdot  \vec{W}_{R\mu})\left(\begin{array}{c} \ell^{c}_i \\ -N^{c}_i\end{array}\right)_L + (\bar{d}^{\prime c}_i,\; -\bar{u}^{\prime c}_{i})_{L}\gamma^{\mu}(g\vec{\tau}\cdot  \vec{W}_{R\mu}-\frac{4}{3}g^{\prime}_{1}B^{\prime}_{1 \mu})\left(\begin{array}{c} d^{\prime c}_i \\ -u^{c}_i\end{array}\right)_L \notag\\
&&-(\bar{\nu},\; \bar{d}^{\prime})_{L}\gamma^{\mu}(g_{X}(\vec{\tau}\cdot  \vec{X}_{\mu})^{\dag})\left(\begin{array}{c} \nu \\ d^{\prime c}\end{array}\right)_L - (\bar{\ell},\; -\bar{u}^{c})_{L}\gamma^{\mu}(g_{X}(\vec{\tau}\cdot  \vec{X}_{\mu})^{\dag})\left(\begin{array}{c} \ell \\ -u_i \end{array}\right)_L\notag \\
&&\left.-(\bar{u}_i,\; \bar{\ell}^{c}_{i})_{L}\gamma^{\mu}(g_{X}(\vec{\tau}\cdot  \vec{X}_{\mu})^{\dag})\left(\begin{array}{c} u_i \\ \ell^{c}_i\end{array}\right)_L - (\bar{d}^{\prime}_i,\; -\bar{N}^{c}_{i})_{L}\gamma^{\mu}(g_{X}(\vec{\tau}\cdot  \vec{X}_{\mu})^{\dag})\left(\begin{array}{c} d^{\prime}_i \\ -N^{c}_i\end{array}\right)_L\right]\label{Lag-Scen-2}.
\end{eqnarray}
In Eq. (\ref{Lag-Scen-2}), the charged part is
\begin{eqnarray}
\mathcal{L}_{\text{int}}(\text{Charged})&=&-\frac{1}{\sqrt{2}}\left[g\left(J_{L}^{+\mu}W_{L \mu}^{+}+J_{R}^{+\mu}W_{R \mu}^{+}+ h.c.\right)-g_{X}\big((\bar{\nu}^{i}\gamma^{\mu}(1-\gamma_{5})d_{i}^{\prime c}-\bar{e}^{i}\gamma^{\mu}(1-\gamma_{5})u_{i}^{\prime c}\right. \notag \\
&&\left.+\bar{u}_{i}\gamma^{\mu}(1-\gamma_{5})e_{i}^{ c}-\bar{d}^{\prime c}_{i}\gamma^{\mu}(1-\gamma_{5})N_{i}^{c})X_{\mu}^{-1/3}+h.c.\big)\right]\label{charged-Lag2},
\end{eqnarray}
whereas the neutral one is
\begin{equation}
\mathcal{L}_{\text{int}} (\text{Neutral})= -g\sin\theta_WJ^{\mu}_{em}A_{\mu}-\frac{g}{\cos{\theta_W}}J_{Z}^{\mu}Z_{\mu}-\frac{g}{\sqrt{1-\tan^{2}}\theta_W}J_{Z^{\prime}}^{\mu}Z^{\prime}_{\mu}-g_{X}\frac{1}{\sqrt{1-g_{1}^2/g^{2}_{X}}}J_{Z^{\prime\prime}}^{\mu}Z^{\prime\prime}_{\mu}.\label{neutral-Lag2}
\end{equation}
Here
\begin{eqnarray}
J^{+ \mu}_{L}&=&\left(\bar{\nu}_{i}\gamma^{\mu}(1-\gamma_{5})e_{i} + \bar{u}_{i}\gamma^{\mu}(1-\gamma_{5})d^{\prime}_{i}\right),\quad\quad\quad J^{+ \mu}_{R}=\left(\bar{N}_{i}\gamma^{\mu}(1+\gamma_{5})e_{i} + \bar{u}_{i}\gamma^{\mu}(1+\gamma_{5})d^{\prime}_{i}\right)\label{current-2},\\
J^{\mu}_{em}&=&\left(-\bar{e}_{i}\gamma^{\mu}e_{i} + \frac{2}{3}\bar{u}_{i}\gamma^{\mu}u_{i}-\frac{1}{3}\bar{d}_{i}\gamma^{\mu}d_{i}\right)\label{current-3},\\
J_{Z}^{\mu} &=&\frac{1}{4}\big[\left(\bar{\nu}_{i}\gamma^{\mu}(1-\gamma_{5})\nu_{i}-\bar{e}_{i}\gamma^{\mu}(1-\gamma_{5})e_{i} +\bar{u}_{i}\gamma^{\mu}(1-\gamma_{5})u_{i} -\bar{d}_{i}\gamma^{\mu}(1-\gamma_{5}) d_{i}\right)-4\sin^{2}\theta_W J^{\mu}_{em},\notag\\
\label{current-4}\\
J_{Z^{\prime}}^{\mu}&=&\frac{1}{4}\big[\tan^{2}{\theta_W}\left(\bar{\nu}_{i}\gamma^{\mu}(1-\gamma_{5})\nu_{i}-\bar{e}_{i}\gamma^{\mu}(1-\gamma_{5})e_{i} +\bar{u}_{i}\gamma^{\mu}(1-\gamma_{5})u_{i} -\bar{d}_{i}\gamma^{\mu}(1-\gamma_{5}) d_{i}-4 J^{\mu}_{em}\right)\notag \\
&&+\left(\bar{N}_{i}\gamma^{\mu}(1+\gamma_{5})N_{i}-\bar{e}_{i}\gamma^{\mu}(1+\gamma_{5})e_{i} +\bar{u}_{i}\gamma^{\mu}(1+\gamma_{5})u_{i} -\bar{d}_{i}\gamma^{\mu}(1+\gamma_{5}) d_{i}\right)\big]
\label{current-5},\\
J_{Z^{\prime\prime}}^{\mu}&=&\frac{1}{4}\big[(\bar{\nu}_{i}\gamma^{\mu}(1-\gamma_{5})\nu_{i}+\bar{N}_{i}\gamma^{\mu}(1+\gamma_{5})N_{i})+2(\bar{e}_{i}\gamma^{\mu}e_{i} +\bar{u}_{i}\gamma^{\mu}u_{i}+\bar{d}_{i}\gamma^{\mu} d_{i})\notag \\
&&-\frac{g^{2}_1}{g^{2}_X}\big(\bar{\nu}_{i}\gamma^{\mu}(1-\gamma_{5})\nu_{i}+\bar{N}_{i}\gamma^{\mu}(1+\gamma_{5})N_{i}+2(-\bar{e}_{i}\gamma^{\mu}e_{i} +\bar{u}_{i}\gamma^{\mu}u_{i}-\bar{d}_{i}\gamma^{\mu} d_{i})\nonumber \\
&&-4(-\bar{e}_{i}\gamma^{\mu}e_{i} +\frac{2}{3}\bar{u}_{i}\gamma^{\mu}u_{i}-\frac{1}{3}\bar{d}_{i}\gamma^{\mu} d_{i})\big)\big]\label{current-6}.
\end{eqnarray}
To obtain the interaction Lagrangian for neutral currents in the final form, we have used the following relations between the neutral vector bosons $W^{0}_{L \mu}, W^{0}_{R \mu}, B_{1 \mu}^{\prime}$, $X^{0}_{\mu}$ and the physical vector bosons $A_{\mu}, Z_{\mu}, Z^{\prime}_{\mu}$, $Z^{\prime\prime}_{\mu}$:
\begin{eqnarray}
gW^{0}_{L \mu}&=& eA_{\mu}+g\cos\theta_{W}Z_{\mu}, \quad\quad\quad gW^{0}_{R \mu} = eA_{\mu}-g\frac{\sin^{2}\theta_W}{\cos\theta_{W}}Z_{\mu}+g\sqrt{1-\tan^{2}\theta_W}Z^{\prime}_{\mu},\notag\\
\label{boson-relations1}\\
g^{\prime}_1B^{\prime}_{1 \mu} &=& eA_{\mu}-g\frac{\sin^{2}\theta_W}{\cos\theta_{W}}Z_{\mu}-g\frac{\tan^{2}\theta_W}{\sqrt{1-\tan^{2}\theta_W}}Z^{\prime}_{\mu}+g_{X}\frac{g^{2}_1}{g^{2}_X}\frac{Z^{\prime \prime}_{\mu}}{\sqrt{1-\frac{g^{2}_1}{g^{2}_X}}}\label{bosn-relations2},\\
g_{X}X^{0}_{\mu} &=& eA_{\mu}-g\frac{\sin^{2}\theta_W}{\cos\theta_{W}}Z_{\mu}-g\frac{\tan^{2}\theta_W}{\sqrt{1-\tan^{2}\theta_W}}Z^{\prime}_{\mu}-g_{X}\sqrt{1-\frac{g^{2}_1}{g^{2}_X}}Z^{\prime \prime}_{\mu}\label{bosn-relations3},\\
g^{\prime}_{1}B^{\prime}_{1\mu}-g_{X}X^{0}_{\mu} &=& g_{X}\sqrt{\frac{1}{1-\frac{g^{2}_1}{g^2_{X}}}} \big[\frac{g^{2}_1}{g^2_{X}}+(1-\frac{g^{2}_1}{g^{2}_X})\big]Z^{\prime\prime}_{\mu} = g_{X}\sqrt{\frac{1}{1-\frac{g^{2}_1}{g^2_{X}}}}Z^{\prime\prime}_{\mu}\label{bosn-relations4}.
\end{eqnarray}
The relations between different couplings are
\begin{eqnarray}
\frac{1}{e^2} &=&\frac{1}{g^2}+\frac{1}{g^{\prime 2}}, \quad\quad\quad \frac{g^{\prime}}{g} = \tan\theta_W,\quad\quad \frac{1}{g^{\prime 2}} =\frac{1}{g^2}+\frac{1}{g_{1}^{2}},\nonumber \\
\frac{1}{g_1^{2}} & = & \frac{1}{g_{1}^{\prime 2}}+\frac{1}{g_{X}^{2}}, \quad\quad\quad\frac{g^{\prime}}{g_{1}} =\sqrt{1-\tan^{2}\theta_W},\nonumber \\
\frac{g}{g^{\prime}_{1}} &=& \sqrt{1-\frac{g^{2}_1}{g^2_{X}}},\quad\quad g^{2}_1 = g^2\frac{\tan^{2}\theta_W}{1-\tan^{2}\theta_W}\label{coupling-relations} .
\end{eqnarray}

The gauge group $\big[SU(2)_{L}\times SU(2)_{R}\times U_{1}(1)_{Y^{\prime}_{1}}\big]\times SU(2)_{X}$ is spontaneously broken to $SU(2)_{L}\times SU(2)_{R}\times U(1)_{Y_{1}}$ by introducing a scalar multiplet $\Sigma$, which belongs to the triplet representation of $SU(2)_X$ but singlet of $SU(2)_{L}\times SU(2)_{R}$ with $Y^{\prime}_{1} = 2/3$. It can be expressed as
\begin{eqnarray}
\Sigma &=&(1, 1, 2/3, 3) = \left(\begin{array}{cc} \Sigma^{1/3}& \Sigma^{2/3} \\
\Sigma^{0} & -\Sigma^{1/3}\end{array}\right)\nonumber \\
& = &\left(\begin{array}{cc} \Sigma^{1/3}& \Sigma^{2/3} \\
\frac{1}{2}(V+H^{0}_{\Sigma}+ih^{0}_{\Sigma}) & -\Sigma^{1/3}\end{array}\right)\rightarrow \left(\begin{array}{cc} 0 & \Sigma^{2/3} \\
\frac{V+H^{0}_{\Sigma}}{2} & 0\end{array}\right),\nonumber \\
\langle \Sigma \rangle & = & \left(\begin{array}{cc} 0 & 0 \\
\frac{V}{2} & 0\end{array}\right) \label{vev-sigma}.
\end{eqnarray}
The scalars $\Sigma^{\pm 1/3}$ and $h^{0}_{\Sigma}$ have been absorbed to give masses to the lepto-quarks $X^{\pm 1/3}_{\mu}$ and the neutral vector boson $Z^{\prime\prime}_{\mu}$, respectively.

The mass term is given by
\begin{eqnarray}
\mathcal{L}_{\text{Mass}} &=& -\frac{1}{8}g^{2}_{X}V^2\big[2X^{1/3 \mu}X^{-1/3}_{\mu}+\frac{2}{g^{2}_X}(g^{\prime}_1 B_{1}^{\prime \mu} - g_{X}X^{0 \mu})(g^{\prime}_1 B_{1 \mu}^{\prime} - g_{X}X_{\mu}^{0})\big]\notag \\
&=& -\frac{1}{8}g^{2}_{X}V^2\left[2X^{1/3 \mu}X^{-1/3}_{\mu}+2\frac{Z^{\prime\prime \mu}Z^{\prime\prime}_{\mu}}{1-g^2_1/g^2_X}\right]\label{mass-term}.
\end{eqnarray}
Hence
\begin{equation}
m^{2}_{X} = \frac{1}{4}g^2_{X}V^2,\quad\quad m^{2}_{Z^{\prime\prime}} = \frac{1}{4}\frac{g_{X}^2 2V^2}{1-g^{2}_1/g^{2}_X} = \frac{2m^{2}_{X}}{1-g^{2}_1/g^{2}_X}\label{mxmass}.
\end{equation}
Thus at the mass scale $m_X$ the lepto-quark $X^{\pm 1/3}$ and the neutral vector boson $Z^{\prime\prime}_{\mu}$ are decoupled and the residual group is the left-right symmetric gauge group $SU(2)_L \times SU(2)_R\times U(1)_{Y}$ that has been extensively studied in the literature.

\subsection{Effective Lagrangian for the LFV decays}

Using Eq. (\ref{charged-Lag2}), the effective Lagrangian relevant for the LFV decays is given by
\begin{eqnarray}
\mathcal{L}_{\text{eff}} &=& \frac{G_X}{\sqrt{2}}\big[\bar{\nu}_{i}\gamma^{\mu}(1-\gamma_5)d^{\prime c}_i - \bar{\ell}_{i}\gamma^{\mu}(1-\gamma_5)u^{c}_i + \bar{u}_{i}\gamma^{\mu}(1-\gamma_5)\ell^{c}_i - \bar{d}^{\prime c}_{i}\gamma^{\mu}(1-\gamma_5)N^{c}_i\big]\times\notag \\
&& \big[\bar{d}^{\prime c}_{j}\gamma_{\mu}(1-\gamma_5)\nu_{j} - \bar{u}^{c}_{j}\gamma_{\mu}(1-\gamma_5)\ell_{j} + \bar{\ell}^{c}_{j}\gamma_{\mu}(1-\gamma_5)u_{j} - \bar{N}^{c}_{j}\gamma_{\mu}(1-\gamma_5)d^{\prime}_{j}\big] \label{Lag-charged}.
\end{eqnarray}
From above equation, the $(\nu_{i}\nu_{j})$ part of the Lagrangian is
\begin{eqnarray}
\mathcal{L}_{\text{eff}}(\nu_{i}\nu_{j}) &=&  \frac{G_X}{\sqrt{2}}\big[\left(\bar{\nu}_{i}\gamma^{\mu}(1-\gamma_5)d^{\prime c}_i\right)\left(\bar{d}^{\prime c}_{j}\gamma_{\mu}(1-\gamma_5)\nu_{j}\right)\big]\notag \\
& = & \frac{G_X}{\sqrt{2}}\big[\left(\bar{d}^{\prime c}_{j}\gamma^{\mu}(1-\gamma_5)d^{\prime c}_i\right)\left(\bar{\nu}_{i}\gamma_{\mu}(1-\gamma_5)\nu_{j}\right)\big]\notag\\
&=&-\frac{G_X}{\sqrt{2}}\big[\left(\bar{d}^{\prime}_{i}\gamma^{\mu}(1+\gamma^5)d^{\prime}_j\right)\left(\bar{\nu}_{i}\gamma_{\mu}(1-\gamma_5)\nu_{j}\right)\big]\label{neutrino-Lag},
\end{eqnarray}
and the part corresponding to $(\ell_{i} \nu_{j})$ is
\begin{eqnarray}
\mathcal{L}_{\text{eff}}(\ell_{i}\nu_{j}) &=&  -\frac{G_X}{\sqrt{2}}\big[\left(\bar{\nu}_{i}\gamma^{\mu}(1-\gamma_5)d^{\prime c}_i\right)\left(\bar{u}^{c}_{j}\gamma_{\mu}(1-\gamma_5)\ell_{j}\right)+h.c.\big]\notag \\
& = & \frac{G_X}{\sqrt{2}}\big[\left(\bar{d}^{\prime}_{i}\gamma^{\mu}(1+\gamma^5)u_{j}\right)\left(\bar{\nu}_{i}\gamma_{\mu}(1-\gamma_5)\ell_{j}\right) + h.c.\big]\label{lnu-Lag}.
\end{eqnarray}
Likewise, the $(\ell_{i} \ell_{j})$ part can be written as
\begin{eqnarray}
\mathcal{L}_{\text{eff}}(\ell_{i}\ell_{j}) &=&  \frac{G_X}{\sqrt{2}}\big[\left(\bar{\ell}_{i}\gamma^{\mu}(1-\gamma^5)u^{c}_i\right)\left(\bar{u}^{c}_{j}\gamma_{\mu}(1-\gamma_5)\ell_{j}\right)\notag\\
&&+\left( \bar{u}_{i}\gamma^{\mu}(1-\gamma^5)\ell^{c}_i\right)\left(\bar{\ell}^{c}_{j}\gamma_{\mu}(1-\gamma_5)u_{j}\right)\big]\notag \\
& = & -\frac{G_X}{\sqrt{2}}\big[\left(\bar{u}_{i}\gamma^{\mu}(1+\gamma^5)u_{j}\right)\left(\bar{\ell}_{i}\gamma_{\mu}(1-\gamma_{5})\ell_{j}\right)\notag\\
&&+\left(\bar{u}_{i}\gamma^{\mu}(1-\gamma^5)u_{j}\right)\left(\bar{\ell}_{i}\gamma_{\mu}(1+\gamma_{5})\ell_{j}\right) \label{ll-Lag}.
\end{eqnarray}

The effective Lagrangian, given in Eq. (\ref{Lag-charged}), is relevant for the LFV decays involving two neutrinos in the final state. The decays of special interest are
\begin{eqnarray}
K^{-} \to \pi^{-}\bar{\nu}_{\mu}\nu_{e}: \quad V_{cs}V^{*}_{ud}; \quad\quad\quad K^{-} \to \pi^{-}\bar{\nu}_{\mu}\nu_{\nu}: \quad V_{cs}V^{*}_{us};\notag\\
\quad\quad K^{-} \to \pi^{-}\bar{\nu}_{e}\nu_{e}: \quad V_{cd}V^{*}_{ud} \label{kpidecays}
\end{eqnarray}
and the decays involving the transition of $b$-quark are
\begin{eqnarray}
&& V_{tb}V^{*}_{cs} \quad\quad\quad\quad\quad\quad V_{tb}V^{*}_{ud} \notag  \\
B^{-} &\to & (K^{-})^{*}\bar{\nu}_{\tau}\nu_\mu , \quad \quad B^{-} \to \rho^{-}(\pi^{-})\bar{\nu}_{\tau}\nu_{e} \notag ,\\
\bar{B}^{0} &\to & (\bar{K}^{0})^{*}\bar{\nu}_{\tau}\nu_\mu  , \quad \quad \bar{B}^{0}\to \pi^{0}\bar{\nu}_{\tau}\nu_e \notag ,\\
\bar{B}_{s}^{0} &\to &\phi \bar{\nu}_{\tau}\nu_\mu  , \quad \quad \quad \quad \bar{B}_{s}^{0}  \to  K^{0}\bar{\nu}_{\tau}\nu_e \notag ,\\
B_{c}^{-} &\to & (D_{s}^{-})^{*}\bar{\nu}_{\tau}\nu_\mu  , \quad \quad B_{c}^{-} \to  D^{-}\bar{\nu}_{\tau}\nu_e \notag,\\
\Lambda_{b} & \to & \Lambda \bar{\nu}_{\tau}\nu_\mu , \quad\quad\quad\quad \bar{B}^{0} \to (\bar{K}^{0})^{*}\bar{\nu}_{\mu}\nu_\mu: V_{ts}V_{cs} \label{38}.
\end{eqnarray}

The effective Lagrangian given in Eq. (\ref{lnu-Lag}), is relevant for the semileptonic LFV decays. The corresponding decays of interest are the following:
\begin{eqnarray}
K^{-}  \to  e^{-}\bar{\nu}_{\mu}, \quad B^{-} \to  e^{-}\bar{\nu}_{\tau}, \quad B_c^{-} \to  \mu^{-}\bar{\nu}_{\tau}, \quad D^{+} \to  \mu^{+}\bar{\nu}_{e}, \quad D_{s}^{+} \to  \mu^{+}\bar{\nu}_{\mu},\notag\\
 \quad D_{s}^{+} \to  \mu^{+}\bar{\nu}_{e}: V_{cd}\;,
K^{-}  \to  \pi^{0} e^{-}\bar{\nu}_{\mu}, \quad \bar{K}^{0}\to \pi^{+} e^{-}\bar{\nu}_{\mu}, \quad D^{*} \to \pi^{0}\mu^{+}\bar{\nu}_{e},\label{decays-1}\\
 \quad D_{s}^{*} \to \phi \mu^{+}\bar{\nu}_{\mu}, \quad D_{s}^{*} \to  K^{* 0}\mu^{+}\bar{\nu}_{e}: V_{cd}\;. \quad
B^{-}  \to  D^{0} e^{-}\bar{\nu}_{\tau}, \label{decays-2}\\ 
\quad \bar{B}^{0}\to D^{+} e^{-}\bar{\nu}_{\tau}, \quad \bar{B}^{0}_{s} \to D^{+}_{s}\mu^{-}\bar{\nu}_{\tau}, \quad B_{c}^{-} \to J/\psi \mu^{-}\bar{\nu}_{\tau}\;,\label{decays-3}\\  
B^{-}  \to  \pi^{0} e^{-}\bar{\nu}_{\tau}, \quad \bar{B}^{0}\to \pi^{+} e^{-}\bar{\nu}_{\tau}, \quad \bar{B}^{0}_{s} \to K^{+}e^{-}\bar{\nu}_{\tau}, \quad B_{c}^{-} \to \bar{D}^{0}e^{-}\bar{\nu}_{\tau}\label{decays-4}.
\end{eqnarray}
Likewise, Eq. (\ref{ll-Lag}) depicts the Lagrangian that is relevant for the LFV decays involving charged leptons. The corresponding decays of interest are
\begin{eqnarray}
D^{0} \to \mu^{+}e^{-}, \quad D^{0} \to \pi^{0}(\rho^{0})\mu^{+}e^{-}, \quad D^{+} \to \pi^{+}(\rho^{+})\mu^{+}e^{-},\notag\\
\quad D_{s}^{+} \to K^{+}\mu^{+}e^{-}, \quad B^{+}_c \to B^{+} \mu^{+}e^{-} , \quad t \to c(u)\tau^{+}\mu^{-}(e^{-}) .\label{decays-5}
\end{eqnarray}

For the case of the gauge group discussed here, the experimental limits on the LFV decays are not very stringent. The experimental value of the branching ratio for the decay $K^{-} \to \pi^{-}\bar{\nu}\nu$ is \cite{pdg}
\begin{equation}
\mathcal{B}(K^{-} \to \pi^{-}\bar{\nu}\nu) = 1.73^{+1.15}_{-1.05}\times 10^{-10} \label{branching-K}.
\end{equation}
The standard model gives \cite{Buras}
\begin{equation}
\mathcal{B}(K^{-} \to \pi^{-}\bar{\nu}\nu) = (8.4 \pm 1.0)\times 10^{-11} \label{SMKnunu}.
\end{equation}

From Eq. (\ref{kpidecays}), we have
\begin{eqnarray}
\mathcal{B}(K^{-} \to \pi^{-}\bar{\nu}_{\mu}\nu_{\mu}) & = & \mathcal{B}(K^{-} \to \pi^{-}\bar{\nu}_{e}\nu_{e}) \notag\\
& = &\left(\frac{G_X}{G_F}\right)^2 2\mathcal{B}(K^{-} \to \pi^{0}e^{-}\bar{\nu}_{e}) = 0.1 \left(\frac{G_X}{G_F}\right)^2\label{kpiLFC-decays},
\end{eqnarray}
\begin{eqnarray}
\mathcal{B}(K^{-} \to \pi^{-}\bar{\nu}_{e}\nu_{\mu}) &=& \mathcal{B}(K^{-} \to \pi^{-}\nu_{e}\bar{\nu}_{\mu})\notag\\
& = &\left(\frac{G_X}{G_F}\right)^2\frac{2}{|V_{us}|^2}\mathcal{B}(K^{-} \to \pi^{0}e^{-}\bar{\nu}_{e})\notag \\
& \approx & (2.0)\times \left(\frac{G_X}{G_F}\right)^2\label{KmBr}.
\end{eqnarray}
By using the experimental value of 
\begin{equation}
\mathcal{B}(K^{-} \to \pi^{0}e^{-}\bar{\nu}_{e}) = (5.07\pm 0.04)\times 10^{-2} \label{ktopi0},
\end{equation}
we get
\begin{equation}
\left(\frac{G_X}{G_F}\right)^2 \leq (0.85 \pm 0.55)\times 10^{-10} \label{GXG-ratio}.
\end{equation}
The other most promising decays that involve the $D$ meson are
\begin{equation}
D^{0} \to \pi^{0} \mu^{+} e^{-}, \quad\quad\quad D^{+} \to \pi^{+} \mu^{+} e^{-}, \quad\quad\quad D_{s}^{+} \to K^{+} \mu^{+} e^{-}.
\end{equation}
For these decays in the limit $m^{2}_{e}/m^2_{\mu} \approx 0$, the differential decay rate becomes
\begin{equation}
\frac{d\Gamma}{dt} = \frac{G_{F}^2}{24\pi^3}K(1-\frac{m_{\mu}^{2}}{t})\left[K^2\left(1+\frac{m^2_{\mu}}{2t}\right)f^2_{+}(t)+\frac{3}{8}\frac{(m^2_{P}-m^2_{P^{\prime}})^2}{m^2_{P}}\frac{m^2_{\mu}}{t}f^2_{0}(t)\right]\label{sLFVD},
\end{equation}
where $m_{P}$ can be $m_{D^0}$, $m_{D^{+}}$ or $m_{D_{s}^{+}}$ and $m_{P^{\prime}}$ can be $m_{\pi^{0}}$, $m_{\pi^{+}}$ and $m_{K^{+}}$. The parameters $K$ and $t$ are defined in Eq. (\ref{sLFV4}) and Eq. (\ref{sLFV5}).

The corresponding semi-leptonic decays in the SM are
\begin{equation}
D^{0} \to \pi^{-} \mu^{+} \nu_{\mu}, \quad\quad\quad D^{+} \to \pi^{0} \mu^{+} \nu_{\mu} ,\quad\quad\quad D_{s}^{+} \to K^{0} \mu^{+} \nu_{\mu}.
\end{equation}
The experimental values of the branching ratios are
\begin{eqnarray}
\mathcal{B}\left(D^{0} \to \pi^{-} \mu^{+} \nu_{\mu}\right) &=& \left(2.37 \pm 0.24\right)\times 10^{-3}\notag .
%\mathcal{B}\left(D^{+} \to \pi^{0} \mu^{+} \nu_e\right) &=& Not existing\notag\\
%\mathcal{B}\left(D_{s}^{+} \to K^{0} \mu^{+} \nu_{\mu}\right) &=& Not existing\notag
\end{eqnarray}
Similarly, the experimental branching ratios for the decay modes involving electron and corresponding neutrino are
\begin{eqnarray}
\mathcal{B}\left(D^{+} \to \pi^{0} e^{+} \nu_{e}\right) &=& \left(4.05 \pm 0.18\right)\times 10^{-3}\notag,\\
\mathcal{B}\left(D_{s}^{+} \to K^{0} e^{+} \nu_{e}\right) &=& \left(3.9 \pm 0.9\right)\times 10^{-3}\notag.\\
\end{eqnarray}

\section{Summary }

The LFV decays are strictly forbidden in the Standard Model. We propose the gauge group $G = SU(2)_L\times U(1)_{Y_1}\times SU(2)_X$ beyond the SM. It provides a framework to derive the effective Hamiltonian for the LFV decays of $K$ and $B$ meson to lepton pairs. The effective coupling constant $\frac{G_{X}}{\sqrt{2}} = \frac{g^{2}_X}{8m^2_{X}}$, where $m_{X}$ is the mass of lepto-quark boson $X_{\mu}^{\pm 2/3}$ corresponding to gauge group $SU(2)_X$ and it gives the mass scale at which the group $G$ is broken to $ SU(2)_L\times U_{Y}(1)$. The lepto-quark $X_{\mu}^{\pm 2/3}$ carry the baryon and lepton numbers, $\Delta B  = 1$ and $\Delta L  = -1$, respectively with $\Delta(B + L) = 0$. The upper bound on the $\frac{G_{X}}{G_{F}}$ is derived from the most stringent experimental limit on $\mathcal{B}(K^{0}_{L} \to \mu^{\mp}e^{\pm}) < 4.7 \times 10^{-12}$. Three cases of paring three generations of leptons and quarks $(i): (1,\; 2,\; 3)$, $(ii): (1,\; 3,\; 2)$ and $(iii): (2,\; 3,\; 1)$ in the representation $(2, \bar{2})$ of $ SU(2)_L\times SU_{X}(2)$ are analyzed. For the assignment $(i)$, the upper bound is derived as  $(\frac{G_{X}}{G_{F}})^2< 1.7 \times 10^{-13}$; while for the assignment $(ii)$ and $(iii)$, the bound is  $(\frac{G_{X}}{G_{F}})^2< 1.1 \times 10^{-10}$. For the experimental observation of the LFV decays of $K$ and $B$ mesons, the assignment $(ii)$ or $(iii)$ is more suitable. In particular for the assignment $(ii)$, $\mathcal{B}(\bar{B}_{s}^{0} \to \mu^{-}\tau^{+}) < 9.5 \times 10^{-10}$, which is promising decay to test our model. In order to get the results for assignment $(iii)$ one has to interchange $\mu$ with $e$.

The decay width for $B^{-} \to K^{-}\mu^{-}\tau^{+}$ is given in Eq. (\ref{sLFV10}) and for $B^{-} \to K^{-}\mu^{-}e^{+}$, in the approximation $m^2_{e}/m_{\mu}^{2}\approx 0$ and in limit $m^2_{\mu}/m^{2}_{B}<< 1$, the differential decay rate involves only one form factor $(f_{+}(t))$. This make it the potential candidate for the experimental detection of the LFV decays. For the assignment $(iii)$ the promising decay channel is  $B^{-} \to K^{-}e^{-}\mu^{+}$.

%its differential decay width can be written as
%\begin{equation}
%\frac{d\Gamma}{dt} = \frac{G_{F}^2}{24\pi^3}|V_{tb}V^{*}_{us}|^2K\left(1-%\frac{m_{\mu}^{2}}{t}\right)^2\left[K^2\left(1+\frac{m^2_{\mu}}{2t}%\right)f^2_{+}(t)+\frac{3}{8}\frac{(m^2_{B}-m^2_{K})^2}{m^2_{B}}%\frac{m^2_{\mu}}{t}f^2_{0}(t)\right].\notag
%\end{equation}
%In the limit $m^2_{\mu}/m^{2}_{B}<< 1$, it involves only one form factor $(f_{+}(t))$. 
%Thus, this is the most promising decay for the experimental detection %of the LFV decays. For the assignment $(iii)$ the promising decay channel is  %$B^{-} \to K^{-}e^{-}\mu^{+}$.

In the second part of this paper, the effective Lagrangian for the LFV decays in the gauge group $\left[SU(2)_{L}\times SU(2)_{R}\times U(1)_{Y_1^{\prime}}\right] \times SU(2)_{X}$ is derived. In this model, the lepto-quark boson $X^{\pm 1/3}_{\mu}$ carry the baryon and lepton number $\Delta B  = -1$ and $\Delta L  = -1$ with $\Delta(B - L) = 0$. The important decays of $K$ and $B$ mesons in this model are depicted in Eqs. (\ref{kpidecays}) and (\ref{decays-5}) along with the $B$-meson decays $B^{-} \to (K^{-})^{*} \bar{\nu}_{\tau}\bar{\nu}_{\mu}$, $\bar{B}_{s}^{0} \to \phi \bar{\nu}_{\tau}\bar{\nu}_{\mu}$, $B^{-}_{c} \to (D^{-}_{s})^{*}\bar{\nu}_{\tau}\bar{\nu}_{\mu}$ and the top-quark decay $t \to c \tau^{+}\mu^{-}$.  From the experimental value of the $\mathcal{B}(K^{-} \to \pi^{-}\nu \bar{\nu}) = (1.7 \pm 1.1)\times 10^{-10}$, one gets $\left(\frac{G_{X}}{G_{F}}\right)^2 < 10^{-10}$. It is hoped that the observation of LFV decays at the LHCb and super B-factories will help us to get constraints on the parameters of the extanded gauge group models that are presented in this work.

\section*{Acknowledgements}
MJA would like to thank the Centre for Future High Energy Physics (CFHEP) at Institute of High Energy Physics (IHEP) for the financial support through visiting scientists scheme. The work of Lu and MJA is partly supported by National  Science Foundation of China (11375208, 11521505, 11621131001 and 11235005).

\section*{Appendix}

In the rest frame of decaying particle $\bar{P}$ for the decay $\bar{P}\to (\bar{P}^{\prime})\ell^{+}\ell^{-}$, we can write
\begin{eqnarray}
m_{P} & = &\omega + E_{1} + E_{2}\notag ,\\
t &=& m^2_{P}-m^2_{P^{\prime}}-2\omega m_{P} = m^2_{1}+ m^2_{2}+2k_{1}\cdot k_2 \label{sLFV4}.
\end{eqnarray}
In above Eq. (\ref{sLFV4}), $\omega$ is the energy of final state meson $\bar{P}^{\prime}$ and $E_{1}$ and $E_{2}$ are the energies of the lepton $\ell_1$ and $\ell_2$, respectively. In order to obtain the expression of the differential decay rate, following relations are useful
\begin{eqnarray}
K^{2}&=&\frac{\left(m^2_{P}-m^2_{P^{\prime}}-t\right)^2-4tm^2_{P^{\prime}}}{2m_{P}} = \frac{\left(m^2_{P}-m^2_{P^{\prime}}+t\right)^2-4tm^2_{P}}{2m_{P}}\notag,\\
E_{1}+E_{2} & = & m_{P}-\omega =\frac{(m^2_{P}-m^2_{P^{\prime}})+t}{2m_{P}} = \sqrt{K^2+t},\label{sLFV5}
\end{eqnarray}
where $K = |\vec{p}^{\prime}|$. 

It is convenient to introduce a four-vector
\begin{eqnarray}
P^{\lambda}& = &(p+p^{\prime})^{\lambda}-\frac{m^2_{P}-m^2_{P^{\prime}}}{t}q^{\lambda}\notag\\
& = &2\left(p^{\lambda}-\frac{m_{P}}{t}(E_{1}+E_{2})q^{\lambda}\right)\label{sLFV6}.
\end{eqnarray}
From above equations, the expression for double diffential decay rate becomes
\begin{equation}
\frac{d^2\Gamma}{dE_{1}dt} = \frac{2G^2}{(2\pi)^3}\frac{1}{8m^2_{P}}\left[f^2_{+}(t)A + f^2_{0}(t)B\right]\label{sLFV7},
\end{equation}
where
\begin{eqnarray}
A &=& 2(P\cdot k_{1})(P\cdot k_2) - (k_{1}\cdot k_2)P^2\notag ,\\
B  &=& \frac{m^2_{P}-m^2_{P^{\prime}}}{t}\left[2(q\cdot k_{1})(q\cdot k_2) - (k_{1}\cdot k_2)q^2\right]\label{sLFV8}.
\end{eqnarray}

\end{document}